\documentclass[12pt]{article}
\usepackage[dvips]{graphicx}
\usepackage[dvips]{graphicx}
\usepackage{graphics}
\usepackage{amssymb}
\usepackage{bm}
\usepackage{amsmath}
\usepackage{color}

\def\spose#1{\hbox to 0pt{#1\hss}}

\def\lta{\mathrel{\spose{\lower 3pt\hbox{$\mathchar"218$}}
     \raise 2.0pt\hbox{$\mathchar"13C$}}}
\def\gta{\mathrel{\spose{\lower 3pt\hbox{$\mathchar"218$}}
     \raise 2.0pt\hbox{$\mathchar"13E$}}}
\newcommand{\be}{\begin{equation}}
\newcommand{\en}{\end{equation}}
\newcommand{\bea}{\begin{eqnarray}}
\newcommand{\ena}{\end{eqnarray}}

\begin{document}

\title{A Lorentz invariant velocity distribution for a relativistic gas}
\author{Evaldo M. F. Curado$^{\mathrm{a,b}}$ and Ivano Dami\~ao Soares$^{\mathrm{a}}$ \\
\emph{   $^{\mathrm{a}}$ Centro Brasileiro de Pesquisas Fisicas   } \\
\emph{   $^{\mathrm{b}}$ Instituto Nacional de Ci\^encia e Tecnologia - Sistemas Complexos}\\
\emph{  Rua Xavier Sigaud 150, 22290-180 - Rio de Janeiro, RJ, Brazil  }
}

\maketitle

\begin{abstract}
We derive a Lorentz invariant distribution of velocities for a relativistic gas. 
Our derivation is based on three pillars: the special theory of relativity, the central limit theorem
and the Lobachevskyian structure of the velocity space of the theory. The rapidity variable
plays a crucial role in our results. For $v^2/c^2 \ll 1$ and $1/\beta=kT/2 m_0 c^2 \ll 1$ the distribution tends to the
Maxwell-Boltzmann distribution. The mean $\langle v^2 \rangle$ evaluated with the Lorentz invariant distribution
is always smaller than the Maxwell-Boltzmann mean and is bounded
by $\langle v^2 \rangle/c^2=1$. This implies that for a given $\langle v^2 \rangle$ the temperature is
larger than the temperature estimated using the Maxwell-Boltzmann distribution. For temperatures of the
order of $T \sim {10^{12}}~ K$ and $T \sim {10^{8}}~ K$ the difference 
is of the order of $10 \%$, respectively for particles with the hydrogen and
the electron rest masses.

\end{abstract}

%\tableofcontents

In physics, it is difficult to overestimate adequately the importance
of the Maxwell-Boltzmann (MB) distribution of velocities
for gases, introduced by Maxwell in 1860 \cite{maxwell1860}. It
was the first time a probability concept was introduced
in a physical theory, as the existing theories at that time,
like Newtonian mechanics and wave theory, were purely
deterministic theories. Actually, the work of Maxwell
was the starting point for Boltzmann to elaborate his research
program, on the evolution of a time dependent distribution of velocities for gases, culminating
in the articles of 1872 \cite{boltzmann1872} and 1877\cite{boltzmann1877}, among
other important papers, which are among the first fundamental
cornerstones of the modern kinetic theory of gases and of
statistical mechanics.
With the implicit use of the atomic theory of matter (at that time a
controversial theory), the new concept of entropy was established
having also as its starting point the introduction, by Maxwell, of the concept of
probability.
Since then the MB distribution played a fundamental role
in the statistical description of gaseous systems with a large number of constituents.
Actually, in many cases it is considerably simpler, and even as accurate as, to use the
MB distribution instead of Bose-Einstein and Fermi-Dirac distributions\cite{balian}.
However a clear limitation of the MB distribution is its nonrelativistic character,
encompassing velocities larger than the velocity of light in contradiction with the special theory of relativity.
\par Our purpose in this paper is to derive a Lorentz invariant distribution of velocities for a relativistic gas 
which has the MB distribution as a limit for velocities which are much smaller than the velocity of light (relatively small temperatures). 
In our derivation we will make use of  an important property connected to the Riemannian structure
of the velocity spaces of Galilean relativity and of Einstein special relativity, namely,
both spaces are maximally symmetric three dimensional (3D) Riemannian spaces, differing only by the fact that in
the Galilean relativity the space is flat (an Euclidean space) while in the Einstein special
relativity the space has a negative constant curvature ($R=-1/c^2$).
This will lead us (i) to use the additivity of velocities in the first case and of rapidities in the latter
case, and (ii) to use Galilean transformations and Lorentz transformations as rigid translations
which map, respectively, each space into itself.
\par We will start by presenting a derivation of the MB distribution which differs from the
derivation used by Maxwell but which is more appropriate for our derivation.
Let us first consider the Galilean addition of velocities in
an Euclidean space. We know that the velocities add according the rule
${\bf v} =  \sum_i {\bf v}_i $. Assuming that the velocities ${\bf v}_i$ are
random variables, with zero mean, and considering that the sum is over
a very large number of particles, we have - by the central limit theorem\cite{central} - that the probability
distribution of velocities for the random variable $\vec{v}$ is given by
$P({\bf v}) \propto \exp(-b{\bf v}^2)$ recovering thus the famous MB distribution.
\par In the special theory of relativity let us consider a fixed inertial reference frame, 
say the laboratory frame. To simplify our discussion we initially assume one dimensional (1D) motion only.
Let the velocity of two particles be $v_1$ and $v_2$, with opposite signs, as measured in this
inertial frame. According to the special theory of relativity the relative velocity of the two particles
must be
{\small
\begin{eqnarray}
\label{eq1}
v=\frac{v_1+v_2}{1+v_1 v_2/c^2}
\end{eqnarray}
}
which is invariant under Lorentz transformations. We note that the addition law (\ref{eq1})
does not satisfy the usual arithmetic addition, as in Galilean relativity, which constitutes
a basic ingredient in the central limit theorem, as is shown above, in our derivation of the
MB distribution. However (\ref{eq1}) can be rewritten as
{\small
\begin{eqnarray}
\label{eq2}
\frac{1-v/c}{1+v/c}=\Big( \frac{1-v_1/c}{1+v_1/c}\Big) \Big(\frac{1-v_2/c}{1+v_2/c}\Big),
\end{eqnarray}
}
and can be extended to any number of particles,
{\small
\begin{eqnarray}
\label{eq3}
\frac{1-v/c}{1+v/c}=\prod_{i}\Big( \frac{1-v_i/c}{1+v_i/c}\Big).
\end{eqnarray}
}
Therefore, if we take the logarithm on both sides (\ref{eq3}) and define
{\small
\begin{eqnarray}
\label{si}
\sigma_i \equiv \frac{1}{2} \ln \left(\frac{1+v_i/c}{1-v_i/c} \right)
= \tanh^{-1}(v_i/c) \,\,,~~\sigma_i \in (-\infty,\infty)
\end{eqnarray}
}
we can express Eq.(\ref{eq3}) as
{\small
\begin{eqnarray}
\label{ssoma}
\sigma= \sum_i \sigma_i            \, .
\end{eqnarray}
}
Let us consider, then, a relativistic gas with a large number of particles.
If we assume that the velocities $v_i$ are independent and random variables, with zero mean,
the variables $\sigma_i$'s are also independent and random
with zero mean, and we have -- in accordance with the central limit theorem -- that the probability distribution of $\sigma$ in an interval $\sigma$ and $\sigma+d\sigma$ is given by,
{\small
\begin{eqnarray}
\label{probs}
P(\sigma)d\sigma= C_1 e^{-\beta \sigma^2} d\sigma \,,
\end{eqnarray}}
Using (\ref{si}) and that $d \sigma = \gamma^2 dv$, where
{\small $\gamma = 1/\sqrt{1-v^2/c^2}$},
we can write the probability distribution for the velocities of a one-dimensional relativistic gas as
{\small
\begin{eqnarray}
\label{eq4}
P(v)dv= C_1 e^{-\beta \Big( \tanh^{-1}(v/c) \Big)^2} \gamma^2 dv,
\end{eqnarray}
}
with the normalization constant $C_1=\frac{1}{c}\sqrt{\beta/\pi}$. The factor {\small $\gamma^2 dv$} 
in (\ref{eq4}) is the invariant line element as shown later.
As we will see the variable $\sigma$ is in fact a particular case of the invariant 
distance measure in a 3D Lobachevsky space, which is the space of velocities
in the special theory of relativity.
\par For the general case of 3D velocities we have that the relative velocity ${\bf v}$ of two particles with
arbitrary velocities ${\bf v_1}$ and ${\bf v_2}$, with respect to a fixed inertial frame, is given by
{\small
\begin{eqnarray}
\label{eq5}
{\bf v}=\frac{{\bf v_1}-{\bf v_2}+(\gamma(v_2)-1)({\bf v_2}/v_2^2)[{\bf v_1}\cdot{\bf v_2}-v_2^2]}{\gamma(v_2)(1-{\bf v_1}\cdot {\bf v_2}/c^2)}.
\end{eqnarray}
}
The above expression also holds with the interchange ${\bf v_1}\leftrightarrows{\bf v_2}$.
The square of the modulus of the relative velocity is given by
{\small
\begin{eqnarray}
\label{eq6}
v^2=\frac{({\bf v_1}-{\bf v_2})^2 - (1/c^2)[{\bf v_1}\wedge{\bf v_2}]^2}{(1-{\bf v_1}\cdot {\bf v_2}/c^2)^2},
\end{eqnarray}
}
which is symmetric with respect to {\small${\bf v_1}$} and {\small ${\bf v_2}$}.
The square of the relative velocity (\ref{eq6})
is invariant under Lorentz transformations. Now following Fock\cite{fock} let us consider
the 3D space of velocities in the special theory of relativity and take two velocities
infinitesimally close, namely, ${\bf v}$ and ${\bf v}+d{\bf v}$.
Let $ds$ be the magnitude of the associated relative velocity divided by $c$. According to (\ref{eq6}) we have
{\small
\begin{eqnarray}
\label{eq7}
d{s}^2=\frac{1}{c^2} \Big( \frac{d{\bf v} \cdot d{\bf v}}{1-v^2/c^2} + \frac{{\bf v} \cdot d{\bf v}}{(1-v^2/c^2)^2} \Big).
\end{eqnarray}
}
Defining ${\bf v}=(v^1,v^2,v^3)$ and introducing the spherical coordinate system in the velocity space,
{\small
\begin{eqnarray}
\label{eq8}
v^1= v \sin \theta \cos \phi,~~v^2= v \sin \theta \sin \phi,~~v^3= v \cos \theta,
\end{eqnarray}
}
we may express (\ref{eq7}) as
{\small
\begin{eqnarray}
\label{eq9}
d {s}^2=\frac{\gamma^4}{c^2} d v^2+ \frac{v^2 \gamma^2}{c^2} d \Omega^2
\end{eqnarray}
}
where $d \Omega^2=d \theta^2 + \sin^2 \theta d \phi^2$.
The determinant of the metric in (\ref{eq9}) is $\sqrt{g}=\gamma^4 v^2 \sin \theta/c^3$ so that the
invariant element of volume of the velocity space in this coordinate system is given by
{\small
\begin{eqnarray}
\label{eq10}
dV=c^3 ~\sqrt{g}~dv d\theta d\phi=\gamma^4 v^2 \sin \theta ~dv d\theta d\phi.
\end{eqnarray}
}
In particular, the invariant line element along an arbitrary direction $\theta=\phi={\rm const. }$ ~is
~$dl=\gamma^2 ~dv$.
The 3D Lobatchevsky velocity space, with the metrics (\ref{eq7}) or (\ref{eq9}), is a maximally symmetric Riemannian space\cite{eisenhart}
and therefore homogeneous, isotropic, with constant Gaussian curvature $R=-1/c^2$ at any point. Since the space is
homogeneous and isotropic any particular velocity may be chosen as the origin. Also, under Lorentz transformations,
the space is mapped rigidly into itself. As a matter of fact the velocity spaces of the
Galilean relativity and of the Einstein special relativity are both 3D maximally symmetric spaces,
differing by the fact that the first has constant curvature $R=0$ and the latter has constant
curvature $R=-1/c^2$. This difference has some bold implications for the relativistic velocity
distribution to be derived. We mention that the remaining possible case of 3D maximally symmetric
space is the spherical space ($R=1/c^2$), which apparently has no physical realization\cite{leblond}.

\par In general we can assert that, since the velocity space is a Lobatchevsky space, the relativistic 
addition theorem for velocity coincides with the vector addition theorem in the Lobatchevsky geometry.
Since the space is homogeneous and isotropic , we can consider that our additive variable is the magnitude of the relative velocity ${\bf v}$, at an arbitrary point taken as the origin. This length is evaluated with
the Lobachevsky metric (11) yielding 
{\small
\begin{eqnarray}
\label{eq11}
s=\pm \tanh^{-1} (|{\bf v}|/c),
\end{eqnarray}
}
which turns out to be the 3D extension of the additive random variable $\sigma=(1/2)\ln \Big(\frac{1+v/c}{1-v/c}\Big)$
used in the derivation of (\ref{eq4}), and which also has zero mean. This Lorentz invariant quantity denotes 
the rapidity of ${\bf v}$ with respect to the origin. The plus or minus sign in (\ref{eq11}) defines the direction of the
rapidity.
\begin{figure}
\begin{center}
\includegraphics*[height=15cm,width=13cm]{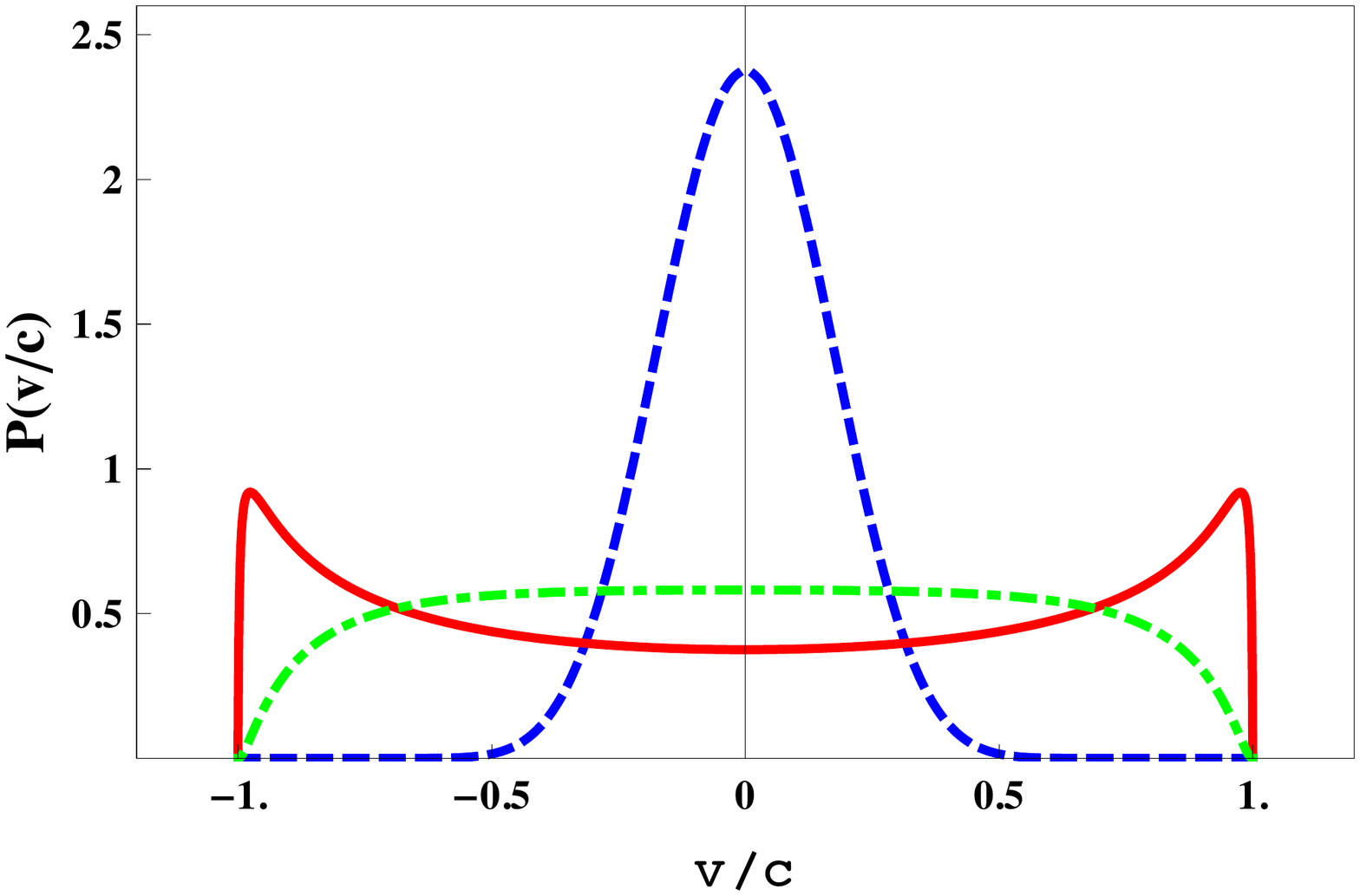}
\caption{Plot of the 1D Lorentz invariant distribution (\ref{eq4})
for a fixed mass and increasing temperatures (successively dashed, dashdotted 
and continuous curves, cf. text). The distribution is zero for $v^2/c^2 \geq 1$,
as expected.}
\label{figdist1}
\end{center}
\end{figure}
\begin{figure}
\begin{center}
\includegraphics*[height=15cm,width=13cm]{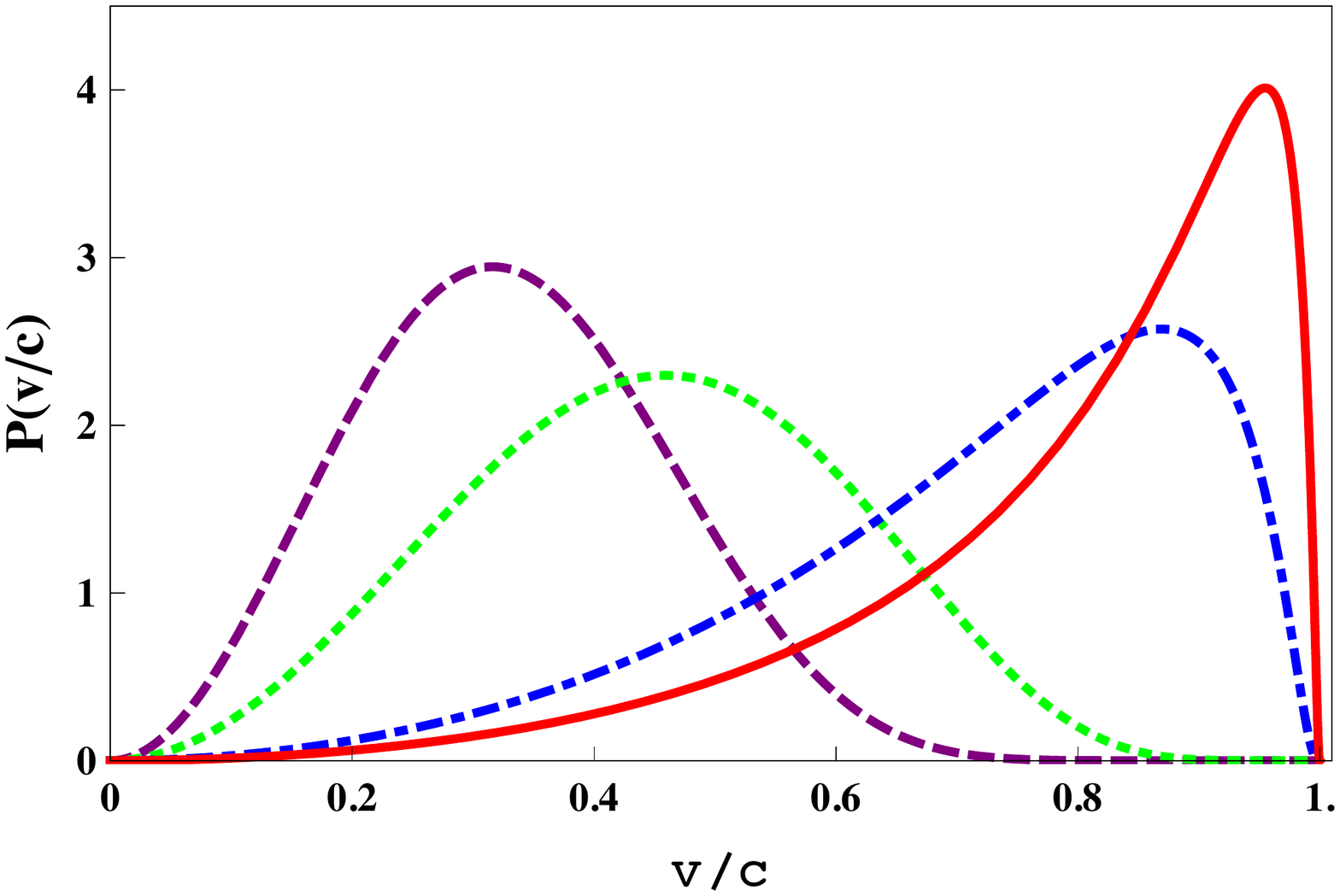}
\caption{Plot of the 3D Lorentz invariant distribution (\ref{eq14})
for increasing temperatures (successively dashed, dotted, dashdotted and 
continuous curves, cf. text). As the temperature
increases the probability piles up towards $v \sim c$ (due to the constraint $v/c \leq 1$),
presenting a delta-type divergence at $v=c$ as $T \rightarrow \infty$.}
\label{figdist2}
\end{center}
\end{figure}
\par With the above ingredients, basically, the homogeneity, isotropy and the invariant volume of the velocity space,
and using the rapidity (\ref{eq11}) as the additive variable, the 1D law (\ref{eq4}) is generalized to
{\small
\begin{eqnarray}
\label{eq13}
P({\bf v})dV= C_3 e^{-\beta \Big( \tanh^{-1}(|{\bf v}|/c) \Big)^2} \gamma^4 dv^1 dv^2 dv^3,
\end{eqnarray}
}
which is the 3D Lorentz invariant distribution law for velocities. In the coordinate system of (\ref{eq9}),
with $dV$ given by (\ref{eq10}), we obtain the expression
{\small
\begin{eqnarray}
\label{eq14}
P(v)dv= 4 \pi C_3 e^{-\beta \Big( \tanh^{-1}(v/c) \Big)^2} \gamma^4 v^2  dv,\\
\nonumber
\end{eqnarray}
}
\noindent where the factor $4 \pi$ comes from the integration in angles. The normalization
constant $C_3$ is given by
{\small
\begin{eqnarray}
\label{eq15}
C_3= \frac{{\sqrt \beta}}{c^3 {\pi}^{3/2} (e^{1/\beta} -1)}.
\end{eqnarray}}
Physically, for one particle, the parameter $\beta$, which must be Lorentz invariant, is taken as the ratio
{\small
\begin{eqnarray}
\label{eq16}
\beta=\frac{m_0 c^2}{2 k T},
\end{eqnarray}
}
where $m_0$ is the rest mass of the particles of the ensemble considered, $T$  the temperature and
$k$ the Boltzmann constant. The nonrelativistic limit corresponds to $v^2/c^2 \ll 1$ and $k~T/m_0 c^2 \ll 1$,
resulting in
{\small
\begin{eqnarray}
\label{eq16b}
(\tanh^{-1}(v/c)~)^2 \simeq v^2/c^2, ~~ C_{3}\simeq \frac{1}{c^3}(\beta/\pi)^{3/2},
\end{eqnarray}
}
which reproduces the MB distribution.

\par Although the components of the two velocities
${\bf v_1}$ and ${\bf v_2}$ appear nonlinearly in Einstein addition law, the relative velocity ${\bf v}$
appearing in (\ref{eq11}) is an invariant and independent variable in the sense that the Lobachevsky space
of velocities is mapped on itself by a Lorentz transformation, this map being a rigid translation without
fixed points, so that no correlation is introduced by a Lorentz transformation. Of course the additivity in
the sense of the Lobachevsky space leads to a distribution which is not separable with respect
to the components of ${\bf v}$.
\begin{figure}
\begin{center}
\vspace{-0.01cm}
\includegraphics*[height=15cm,width=13cm]{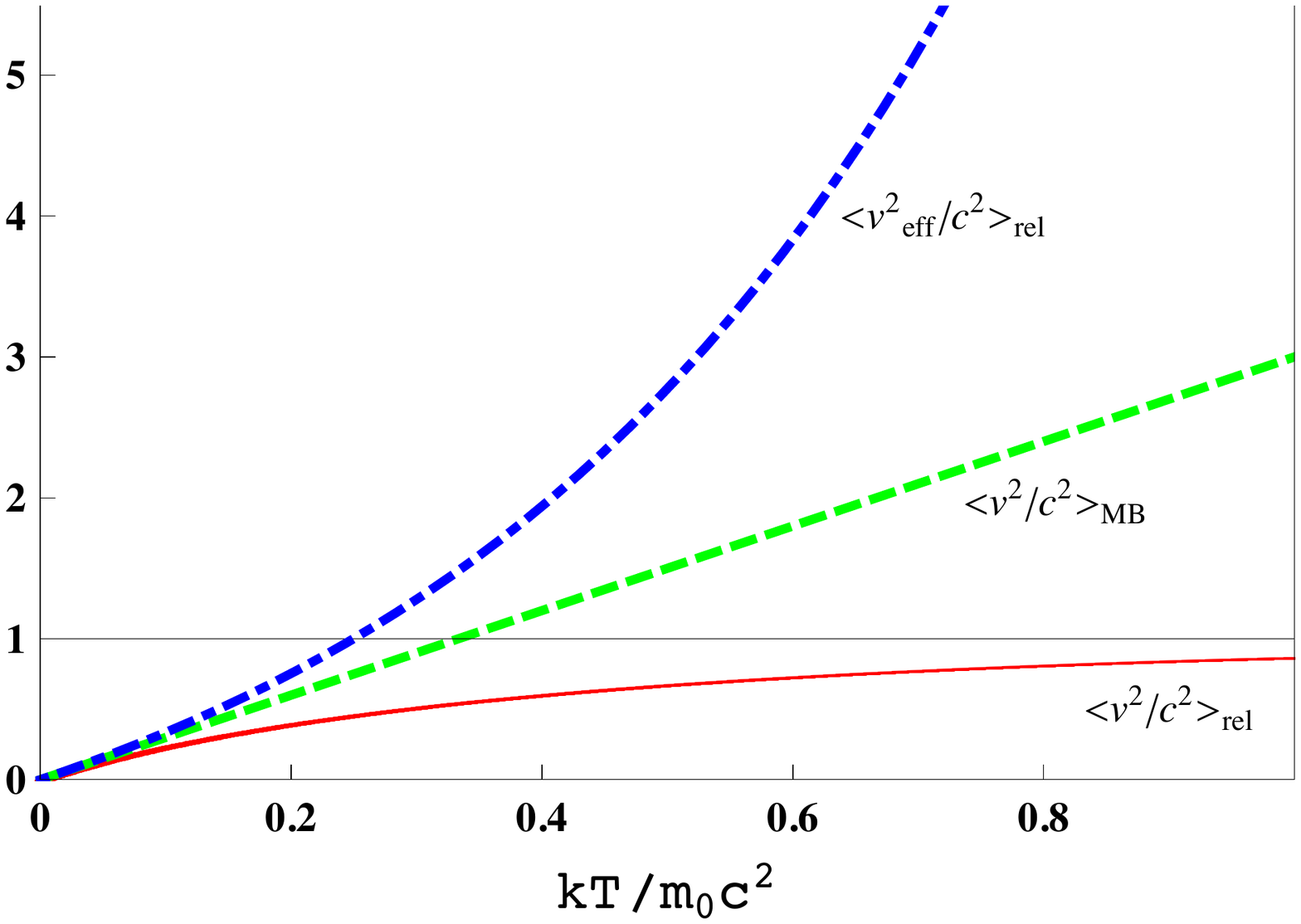}
\caption{The means of $v^2/c^2$ using the MB (dashed curve) and the Lorentz invariant (continuous curve) 3D distributions.
The mean $\langle v^2/c^2\rangle_{\rm rel}$ is asymptotically limited by $1$.
The dashdotted curve is the mean $\langle v_{\rm eff}^{2}/c^2 \rangle_{\rm rel}$ of the effective relativistic velocity (\ref{eq18})
using the Lorentz invariant distribution.}
\label{fig3}
\end{center}
\end{figure}
{\small
\begin{center}
{\small
\begin{table*}
\caption{~~Comparing the velocity means evaluated with the Lorentz invariant and the MB distributions.
The difference between the mean relativistic kinetic energy and the MB kinetic energy, $\Delta$,
corresponds also to the discrepancy of the masses of virialized gravitational systems estimated with both distributions, for a fixed gravitational radius.}
\vspace{0.15cm}
\begin{tabular}{|c|c|c|c|c|}
\hline
 %&  &  &  &   \\
$kT/m_0 c^2$ & $\langle v_{\rm eff}^{2} \rangle_{\rm rel}/c^2 $& $\langle v^{2} \rangle_{\rm rel} /c^2$ &${\langle v^{2} \rangle}_{MB}/c^2$ &$\Delta=(\langle v_{\rm eff}^{2} \rangle_{\rm rel}-{\langle v^{2} \rangle}_{MB})/ \langle v_{\rm eff}^{2} \rangle_{\rm rel}$ \\
 %&  &  &  &    \\
\hline
% &  &  &  &   \\
$0.01$ & $0.03032 $ &$0.02922$ &$ 0.03$ &$ 1.08~\%$ \\
%&  &  &  &   \\
\hline
%&  &  &  &   \\
$0.10$ & $0.33529 $ &$0.23927$ &$ 0.30$ &$ 10.52~\%$ \\
%&  &  &  &   \\
\hline
%&  &  &  &   \\
$0.25$ & $1.00139 $ &$0.46505$ &$ 0.75$ &$ 25.10~\%$ \\
%&  &  &  &   \\
\hline
%&  &  &  &   \\
$0.50$ & $2.77437 $ &$0.68145$ &$ 1.50$ &$ 45.93~\%$ \\
%&  &  &  &   \\
\hline
%&  &  &  &   \\
$1.00$ & $11.83122 $ &$0.87657$ &$ 3.00$ &$ 74.64~\%$ \\
%&  &  &  &   \\
\hline
\end{tabular}
\end{table*}}
\end{center}}
\par In Figs. \ref{figdist1} and \ref{figdist2} we illustrate the behavior of the 1D and 3D
Lorentz invariant distributions. Contrary to the case of the nonrelativistic
distributions, which extends to infinite velocities, the Lorentz invariant distributions are
bounded to $v^2/c^2 \leq 1$. As a consequence we see that in the relativistic case,
as the temperature increases (or $\beta$ decreases) we have a piling up of
the probability towards $v^2/c^2 \sim 1$, diverging at $v^2/c^2=1$ as $\beta \rightarrow 0$.
In both figures we adopted $m_0$ as the hydrogen mass. The temperatures used in Fig. \ref{figdist1} were
$T=3 \times 10^{11},~5 \times 10^{12} ~{\rm and}~ 1.2 \times 10^{13}~K$ while in Fig. \ref{figdist2},
$T=5 \times 10^{11},~1 \times 10^{12},~3.5 \times 10^{12} ~{\rm and}~ 5 \times 10^{12}~K$.
\par Using the Lorentz invariant distribution (\ref{eq14}) we now evaluate the mean of $v^2/c^2$,
{\small
\begin{eqnarray}
\label{eq17}
\langle v^2 \rangle_{\rm rel}/c^2= 4 \pi  c^3 C_3 \int^{\infty}_{0} \exp(-\beta x^2)\tanh^{2} x~ \sinh^{2} x~ dx.
\end{eqnarray}
}
In Fig. \ref{fig3} we compare $\langle v^2 \rangle_{\rm rel}/c^2$  with the MB {\small ${\langle v^2 \rangle}_{MB}=3 k T/m_0$}.
We see that $\langle v^2 \rangle_{\rm rel}/c^2$ is asymptotically limited by $1$,
in contrast with the straight line corresponding to the MB mean.
As a consequence, when the Lorentz invariant distribution is considered,
a given $\langle v^2 \rangle_{\rm rel}$ corresponds to a temperature which is higher than the temperature
obtained with the MB mean. We note that a discrepancy of the order of $10\%$ between the two velocity means 
corresponds to a temperature $T \sim {10^{12}}~K$ for particles with the hydrogen mass.
\par For the relativistic kinetic energy {\small $E_k= E-m_0 c^2$}, which includes higher
order contributions in the velocity, we define an effective velocity $v_{\rm eff}$ by
{\small
\begin{eqnarray}
\label{eq18}
v_{\rm eff}^{2} \equiv \frac{2 E_k}{m_0}= 2 c^2 \Big(\gamma-1 \Big)=v^2 \Big(1+ \frac{3}{4}~\frac{v^2}{c^2}+... \Big),
\end{eqnarray}
}
whose mean in the Lorentz invariant distribution is
{\small
\begin{eqnarray}
\label{eq19}
{\langle v_{\rm eff}^{2} \rangle_{\rm rel}}/c^2 = 2\Big(\frac{e^{1/4\beta} \sinh (1/\beta)}{1-e^{-1/\beta}}-1 \Big).
\end{eqnarray}
}
The mean (\ref{eq19}) has the MB mean as a lower bound, 
as illustrated in Fig. \ref{fig3}. Actually (\ref{eq19}) yields the relativistic
extension for the mean kinetic energy, 
{\small
\begin{eqnarray}
\label{eq20}
\langle E_k \rangle_{\rm rel}=\frac{1}{2} m_0 \langle v_{\rm eff}^{2} \rangle_{\rm rel} = m_0 c^2\Big(\frac{e^{1/4\beta} \sinh (1/\beta)}{1-e^{-1/\beta}}-1 \Big).
\end{eqnarray}
}
For large $\beta$, the right hand side of (\ref{eq19}) yields the MB limit
$\langle v_{\rm eff}^{2} \rangle \simeq 3 k T/m_0$, as expected. The exact equation (\ref{eq20})
corrects the mean relativistic energy expression given in Ref. \cite{tolman}, where the MB distribution was used.
\par Some comments are in order now. As mentioned before, the mean $\langle v^{2} \rangle_{\rm rel}$ is associated with a large
temperature as compared to the MB mean, therefore leading to a even higher relativistic kinetic energy
than the one obtained with the Maxwell-Boltzmann distribution. This has an important consequence
for virialized gravitational systems since, for a fixed temperature and gravitational radius, the mass of the virialized system
results larger for the Lorentz invariant distribution than for the MB distribution, as illustrated in Table I. 
For a difference $\Delta$ of the order of $10\%$, which might lead to substantial
observable effects, we obtain temperatures $T \simeq 1.062 \times 10^{12}~K$ and $T \simeq 5.93 \times 10^{8}~K$,
respectively for hydrogen and electrons in thermal equilibrium. Such high temperatures can possibly occur in
astrophysical scenarios. High temperatures presently observed in astrophysical systems, say $T \sim {10^{8}}~K$, 
yield discrepancies $\Delta \simeq 0.001\%$ and $\Delta \simeq 1.81\%$, respectively for hydrogen and electrons.
A distinct scenario where the relativistic effects could be dominant are quark-gluon plasma formed in ultrarelativistic
nucleus-nucleus collisions, under the assumption of local thermal equilibrium. The temperatures involved in these
systems are of the order of $10^{15}~K$\cite{bjorken,adcox}. 
\par  Finally a corollary following from our derivation (\ref{eq14}) is the
behavior of the temperature under a Lorentz transformation. Since the $\beta$ parameter (\ref{eq16})
must be a invariant under a Lorentz transformation two outcomes are possible, as we are assuming that the
Boltzmann constant is a relativistic invariant. Either (i) we consider $m_0 c^2 \equiv E^2/c^2 -p^2$ the invariant four-momentum
norm, implying that the temperature is also invariant, or (ii) $m_0$ is the rest energy implying then that $T \rightarrow T^{\prime}=\gamma T$.
We favor the option (i) in accord with the considerations of Landsberg \cite{landsberg}.
\par We are grateful to Prof. Constantino Tsallis for the stimulating
discussions and suggestions, which were fundamental to the development of this paper. We also
acknowledge the Brazilian scientific agencies CNPq, FAPERJ and CAPES for financial support.

\end{document}